\documentclass[iop,apj,12pt,twocolappendix]{emulateapj}

\usepackage{acronym}
\usepackage{graphicx}
\usepackage{amsmath,amsfonts,amssymb}
\usepackage{mathrsfs}
\usepackage[utf8]{inputenc}
\usepackage[normalem]{ulem}

\usepackage[plainpages=false, colorlinks=true, anchorcolor=blue, linkcolor=blue, citecolor=blue, bookmarks=false]{hyperref}
\usepackage{natbib}
\epsscale{1.1}

\usepackage{tensind}
\tensordelimiter{?}
\tensorformat{lrc}

\newcommand{\code}[1]{\texttt{#1}}

\newcommand{\eg}{\textit{e.g.}}

\newcommand{\run}[1]{\mbox{$\ell_{\rm mix}$-{#1}-m}}

\begin{document}

\title{General-Relativistic Large-Eddy Simulations of Binary Neutron
Star Mergers}
\author{David Radice}
\affiliation{Institute for Advanced Study,
1 Einstein Drive, Princeton, NJ 08540, USA}
\affiliation{Department of Astrophysical Sciences, Princeton University,
4 Ivy Lane, Princeton, NJ 08544, USA}

\begin{abstract}
The flow inside remnants of binary neutron star (NS) mergers is expected
to be turbulent, because of magnetohydrodynamics instability activated
at scales too small to be resolved in simulations. To study the
large-scale impact of these instabilities, we develop a new formalism,
based on the large-eddy simulation technique, for the modeling of
subgrid-scale turbulent transport in general relativity. We apply it,
for the first time, to the simulation of the late-inspiral and merger of
two NSs. We find that turbulence can significantly affect the structure
and survival time of the merger remnant, as well as its
gravitational-wave (GW) and neutrino emissions. The former will be
relevant for GW observation of merging neutron stars. The latter will
affect the composition of the outflow driven by the merger and might
influence its nucleosynthetic yields. The accretion rate after
black-hole formation is also affected. Nevertheless, we find that, for
the most likely values of the turbulence mixing efficiency, these
effects are relatively small and the GW signal will be affected only
weakly by the turbulence. Thus, our simulations provide a first
validation of all existing post-merger GW models.
\end{abstract}



\section{Introduction} %
\label{sec:intro}
The typical outcome of the merger of two \acp{NS} is expected to be the
formation of an \ac{HMNS}: a massive \ac{NS} temporarily supported
against gravitational collapse by its fast differential rotation,
although prompt \ac{BH} formation might occur for large masses and/or
soft \acp{EOS} \citep[and references therein]{baiotti:2016qnr}. Its
survival time and, in general, its properties, are important for the
multimessenger signature of \ac{NS} mergers and for their
nucleosynthetic yields.  The \ac{HMNS} has a magnetar-level B-field, and
it is a bright source of neutrinos \citep{sekiguchi:2011zd,
kiuchi:2014hja}. These could drive baryon-rich winds
\citep{dessart:2008zd, siegel:2014ita}. The presence of an \ac{HMNS}
could significantly boost the neutrino annihilation rates at
high-latitudes \citep{richers:2015lma, perego:2017fho} and perhaps
contribute to the launching of a relativistic jet and a \ac{SGRB}
\citep{nakar:2007yr}. Neutrinos could also affect the yield and
electromagnetic signature of the r-process nucleosynthesis in the binary
ejecta \citep{wanajo:2014wha, metzger:2014ila}. Long-lived massive
\acp{NS} created in mergers might power the X-ray tails observed in some
\ac{SGRB} \citep{rowlinson:2013ue, lasky:2013yaa, gao:2015xle}.
Finally, \acp{GW} from the \ac{HMNS} could be used to constrain its
\ac{EOS} \citep{bauswein:2011tp, takami:2014zpa, bernuzzi:2015rla,
radice:2016rys}.

Despite the rapid recent progress of \ac{GRMHD} \acused{GR} simulations
\citep{rezzolla:2011da, kiuchi:2014hja, ruiz:2016rai}, the impact of
magnetoturbulence on the structure and survival time of the \ac{HMNS} is
highly uncertain. The \ac{MRI} \citep{balbus:1991ay} is expected to
operate inside the \ac{HMNS}, drive the redistribution of angular
momentum and affect its lifetime and properties \citep{duez:2006qe,
siegel:2013nrw}. Unfortunately, the fastest growing mode of the \ac{MRI}
in these systems is inaccessible even to the highest-resolution
simulations \citep{kiuchi:2015sga}.

A possible way to model the impact of turbulent transport of angular
momentum in the \ac{HMNS} would be to use an effective viscosity
\citep{duez:2004nf}. This approach is made difficult by the fact that
the Navier-Stokes equations describing relativistic viscous flows are
known to exhibit a number of unphysical pathologies
\citep{hiscock:1985zz, kostadt:2000ty}. There are more complex fluid
models that do not have these shortcomings \citep{andersson:2006nr,
rezzolla:2013book}. However, they are also not entirely without problems
\citep{majorana:1985a, hiscock:1988jd}, are difficult to implement
\citep[\eg][]{takamoto:2011wi}, and their non-linear properties are
poorly understood. More importantly, they contain a large number of
transport coefficients that have no classical counterpart.  These have
no clear physical meaning and are not even in principle measurable
\citep{geroch:1995bx, lindblom:1995gp}.

Here, we propose an alternative approach. Our starting point is the
observation that turbulence models do not have to be restricted to the
class of equations describing fluids with physical viscosity or heat
transfer. Instead, we develop an effective model based on a \ac{GR}
extension of the Newtonian \ac{LES}\acused{GRLES} framework
\citep[\eg][]{miesch:2015a}. Our model, while recovering the Navier-Stokes
equations in the Newtonian limit, does not correspond to or have the
same limitations as any relativistic theory of viscous flows.

In this \emph{Letter}, after a brief description of the \ac{GRLES}
formulation, we present, for the first time, simulations in full-\ac{GR}
of merging \ac{NS} with a realistic, tabulated, nuclear \ac{EOS}, neutrino
cooling, and parametrized turbulent transport. We show that turbulence
could influence the structure of the \ac{HMNS}, as well as its \ac{GW}
and neutrino emissions. On the other hand, for the most realistic values
of the turbulent viscosity, these effects appear to be small and our
simulations provide an important confirmation of a number of previous
results where turbulent transport was not included.

\section{Formulation} %
\label{sec:formulation}
Our starting point is the stress energy tensor of a perfect fluid
\begin{equation}\label{eq:tmunu}
  ?[c]T_\mu_\nu? = \rho h u_\mu u_\nu + p ?[c]g_\mu_\nu?\,,
\end{equation}
where $\rho$, $h$, $u_\mu$ and $g_{\mu\nu}$ are, respectively, density,
specific enthalpy, four-velocity, and the spacetime metric.

In numerical relativity, spacetime is decomposed in space-like slices
with normal $n^\mu$. We decompose $?[c]T_\mu_\nu?$ with respect to
$n^\mu$ as
\begin{equation}\label{eq:tmunu.decomp}
  ?[c]T_\mu_\nu? = E n_\mu n_\nu + S_{\mu} n_{\nu} + S_{\nu} n_{\mu}  +
  S_{\mu\nu}\,,
\end{equation}
where
\begin{align}
  &E = T_{\mu\nu} n^\mu n^\nu = \rho h W^2 -p\,,  \\
  &S_\mu = - \gamma_{\mu\alpha} n_\beta T^{\alpha\beta} = \rho h W^2
  v_\mu\,, \\
  &S_{\mu\nu} = \gamma_{\mu\alpha} \gamma_{\mu\beta} T^{\alpha\beta}
  = S_\mu v_\nu + p \gamma_{\mu\nu}\,,
\end{align}
and $\gamma_{\mu\nu}$, $v^\mu$, $p$, and $W$ are, respectively, the
spatial metric, the three-velocity, the pressure, and the Lorentz
factor.

The equations of energy and momentum conservation are
\begin{eqnarray}
&&\begin{split}\label{eq:momentum}
  \partial_t \big(\sqrt{\gamma} & S_i\big)  + \partial_j \Big[
    \alpha\sqrt{\gamma}\big( ?[c]S_i^j? + S_i n^j  \big)
    \Big] = \\
  &\alpha\sqrt{\gamma}\Big(\frac{1}{2} S^{jk} \partial_i \gamma_{jk} +
  \frac{1}{\alpha} S_k \partial_i \beta^k - E \partial_i \log \alpha
  \Big)\,,
\end{split} \\
&&\begin{split}\label{eq:energy}
  \partial_t \big(\sqrt{\gamma} E\big) +& \partial_j \Big[
    \alpha\sqrt{\gamma}\big(S^j + E n^j\big)\Big] = \\
  &\alpha\sqrt{\gamma} \Big(K_{ij} S^{ij} - S^i \partial_i \log \alpha
    \Big)\,,
\end{split}
\end{eqnarray}
where $\alpha$, $\beta^i$, $K_{ij}$ are, respectively, the lapse
function, shift vector, three-metric, and extrinsic curvature.
$\sqrt{\gamma}$ is the spatial volume element. These equations are then
closed with an \ac{EOS} and equations describing the conservation of the
baryon and lepton numbers.

Equations~\eqref{eq:momentum} and \eqref{eq:energy} contain modes at all
scales, but, in numerical simulations, only modes resolved with
sufficiently many grid zones can develop. In essence, any simulation
deals only with a coarse-grained version of the hydrodynamics equations.
In the \ac{LES} framework, this observation is made rigorous with the
introduction of a linear filtering operation $u\mapsto\overline{u}$ that
removes features at scales smaller than a given $\Delta$. Here, we adopt
for the filtering operator the cell-averaging of the finite-volume
discretization of the equations. We leave the investigation of more
advanced filters for future work. If we filter Eqs.~\eqref{eq:momentum}
and \eqref{eq:energy} we obtain
\begin{eqnarray}
&&\begin{split}\label{eq:momentum.filter}
  \partial_t \big(\sqrt{\gamma}  &\overline{S_i}\big) + \partial_j \Big[
    \alpha\sqrt{\gamma}\big( \overline{?[c]S_i^j?} + \overline{S_i} n^j  \big)
    \Big] =\\
   &\alpha\sqrt{\gamma}\Big(\frac{1}{2} \overline{S^{jk}} \partial_i
   \gamma_{jk} + \frac{1}{\alpha} \overline{S_k} \partial_i \beta^k -
   \overline{E} \partial_i \log \alpha \Big)\,,
\end{split} \\
&&\begin{split}\label{eq:energy.filter}
  \partial_t \big(\sqrt{\gamma} \overline{E}\big) +& \partial_j \Big[
    \alpha\sqrt{\gamma}\big(\overline{S^j} + \overline{E} n^j\big)\Big]
    = \\
  &\alpha\sqrt{\gamma} \Big(K_{ij} \overline{S^{ij}} - \overline{S^i}
  \partial_i \log \alpha \Big)\,.
\end{split}
\end{eqnarray}

Note that Eqs.~\eqref{eq:momentum.filter} and \eqref{eq:energy.filter}
are exact, but are not closed. The reason is that $\overline{S_i v_j}$
cannot be expressed only in terms of other coarse-grained quantities. A
closure is needed:
\begin{equation}
  \overline{S_i v_j} = \overline{S_i} \overline{v_j} + \tau_{ij}\,.
\end{equation}
$\tau_{ij}$ is the so-called subgrid-scale turbulent tensor.  Similar
terms appear in the coarse graining of the baryon and lepton number
conservation equations, but, for simplicity, we will neglect them here.
Simulations usually assume $\tau_{ij} = 0$. Here, instead, we will use
$\tau_{ij}$ to model small-scale turbulence in merger simulations. To do
so, in analogy with the classical Newtonian closure of
\citet{smagorinsky:1963a}, we choose the ansatz
\begin{equation}\label{eq:turb.visc.1}
  \tau_{ij} = - 2 \nu_T \rho h W^2 \left[ \frac{1}{2} \big(\nabla_i
  \overline{v_j} + \nabla_j \overline{v_i}\big) - \frac{1}{3} \nabla_k
  \overline{v^k} \gamma_{ij} \right]\,,
\end{equation}
where $\nabla$ is the covariant derivative compatible with
$\gamma_{ij}$.  The quantity $\nu_T$ has a dimension of a viscosity. On
dimensional grounds, we are led to assume
\begin{equation}\label{eq:turb.visc.2}
  \nu_T = \ell_{\rm mix} c_s\,,
\end{equation}
where $c_s$ is the local sound speed, and $\ell_{\rm mix}$, often called
the mixing length, is a characteristic length over which turbulence operates.
Note that $\nu_T$ is not a physical viscosity; indeed, its definition
depends on the numerical grid and on the Eulerian observer $n^\mu$. This
is expected, because the notion of resolved and unresolved scales is
observer dependent in relativity. $\nu_T$ should be calibrated on the
basis of highly-resolved simulations and/or using self-similarity
methods \citep[\eg][]{germano:1991a}. We leave this task for future
work. For now, we will treat $\ell_{\rm mix}$ as a free parameter.
Assuming \ac{MRI} turbulence, it is then natural to set $\ell_{\rm mix}
\sim \lambda_{\rm MRI}$, where \citep{duez:2006qe}
\begin{equation}\label{eq:lambdamri}
  \lambda_{\rm MRI} \sim 3\ {\rm m}\ \left(\frac{\Omega}{4\ {\rm rad}\ {\rm
  ms}^{-1}}\right)^{-1} \left(\frac{B}{10^{14}\ {\rm G}}\right)\,.
\end{equation}

Equations~\eqref{eq:momentum.filter}, \eqref{eq:energy.filter},
~\eqref{eq:turb.visc.1}, and \eqref{eq:turb.visc.2}, together with the
\ac{EOS}, and the continuity equations are what we refer to as the
\ac{GRLES} equations. We verified, by repeating the
analysis of \citet{hiscock:1985zz} and numerically, that the \ac{GRLES}
equations are not affected by the same pathologies as the relativistic
Navier-Stokes equations.

\section{Implementation} %
\label{sec:implementation}

\begin{figure}
  \plotone{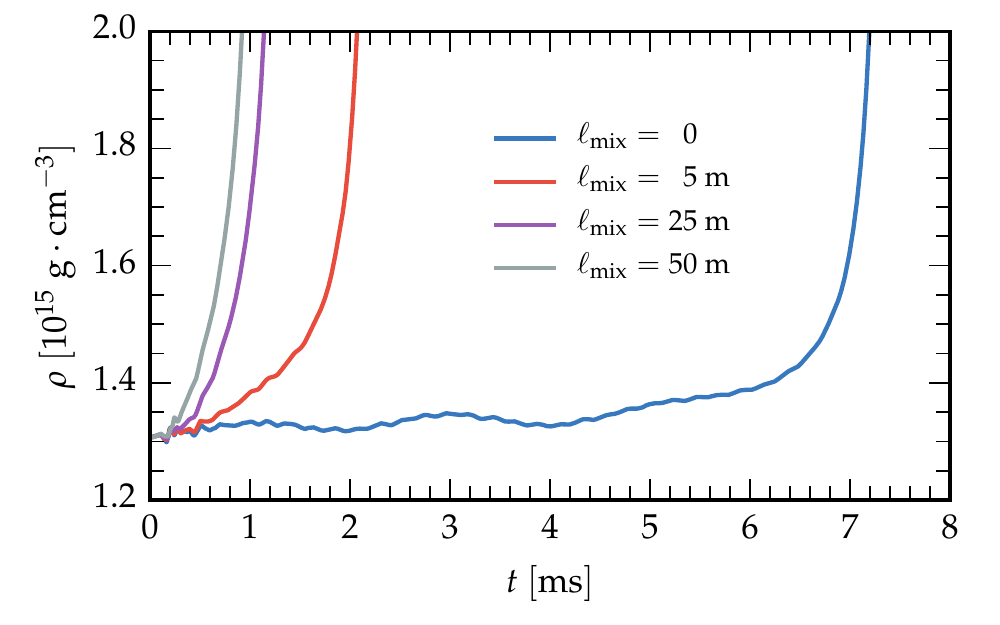}
  \caption{Maximum density in the collapse of a differentially rotating
  equilibrium configuration. Turbulent transport of angular momentum
  leads to an accelerated collapse.}
  \label{fig:rns.collapse}
\end{figure}

We implement the \ac{GRLES} equations into the \ac{GRHD} code
\code{WhiskyTHC} \citep{radice:2012cu, radice:2013hxh, radice:2013xpa}.
With our current choice of the filtering operator, this only amounts to
the inclusion of $\tau_{ij}$ in the equations. We treat the viscous
fluxes in a flux-conservative way and we self-consistently include the
turbulent stress-tensor in the energy and momentum source terms, as well
as in the calculation of the spacetime geometry.

For the simulations presented here, we use the microphysical \ac{EOS} of
\citet{lattimer:1991nc} with nuclear compressibility parameter $K = 220\
{\rm MeV}$. Neutrino cooling is treated with the scheme presented in
\citep{radice:2016dwd}. For $\ell_{\rm mix}$, we consider the values 0
(our reference run), 5, 25, and 50 meters. Over this range, 5 meters is
the most likely value for $\ell_{\rm mix}$ given
Eq.~\eqref{eq:lambdamri}, while 50 meters might be unphysically large,
in the light of the lack of convergence observed in the 17-meter
resolution simulation of \citet{kiuchi:2015sga}.

As a first example, we consider the evolution of an equilibrium
configuration constructed with the \code{RNS} code
\citep{stergioulas:1994ea}. The initial configuration has gravitational
mass $M \simeq 2.45\ M_\odot$ and angular momentum $J/M^2 \simeq 0.66\
G/c$. We use the differential rotation law of \cite{komatsu:1989a},
which, in the Newtonian limits reduces to
\begin{equation}\label{eq:jconst}
  \Omega = \frac{\Omega_c}{1 + \left(\frac{\varpi}{R_e}\right)^2}\,,
\end{equation}
where $\varpi$ is the cylindrical radius, $\Omega_c$ is the angular
velocity at the center, and $R_e$ is the stellar equatorial radius. The
resolution for this test is $\simeq 370\ {\rm m}$.

We plot the maximum density as a function of time in
Fig.~\ref{fig:rns.collapse}. As expected on the basis of previous work
\citep{duez:2004nf}, the inclusion of turbulent viscosity results in the
transport of angular momentum leading to gravitational collapse.  This
test shows that \code{WhiskyTHC} is able to capture the effect of
turbulent viscosity even at low resolution.

\section{Binary Neutron Star Mergers} %
\label{sec:results}
We consider the last ${\sim}4$ orbits and merger of two $1.35$-$M_\odot$
\acp{NS}. We already evolved this binary in \citet{bernuzzi:2015opx},
where a description of the properties of the initial data is also given.
For the evolution, we use the high-resolution setup of
\citet{bernuzzi:2015opx}, with the improvements discussed in
\citet{radice:2016rys}. We perform simulations with resolutions, on the
finest refinement level of ${\sim} 185\ {\rm m}$ and ${\sim} 246\ {\rm
m}$. We present results from the high-resolution simulations. In the
low-resolution simulations, there are quantitative, but not qualitative
differences.

\begin{figure*}
  \plottwo{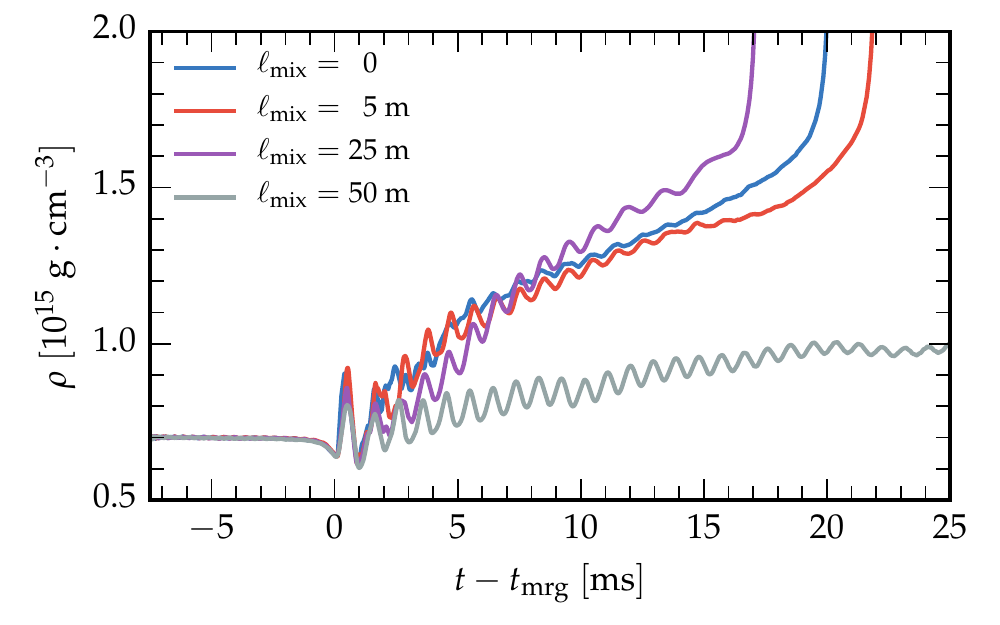}{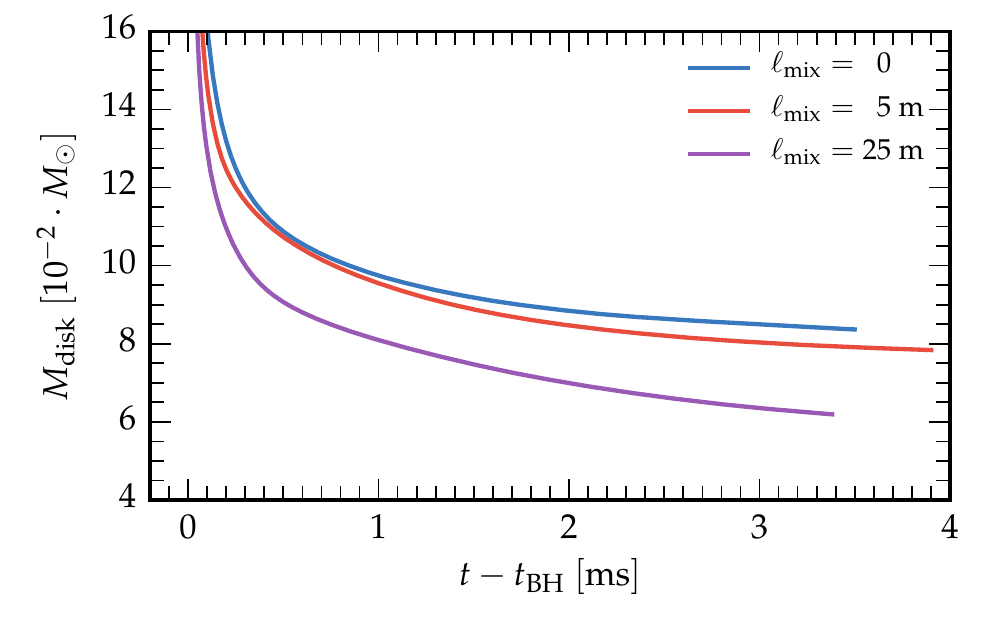}
  \caption{Maximum density (\emph{left panel}) and (baryonic) disk mass
  (\emph{right panel}). The disk mass is computed as the total mass
  outside the apparent horizon. The impact of turbulent mixing on the
  compactness of the \ac{HMNS} is non-trivial and non-monotonic.
  Turbulent angular momentum transport results in larger accretion rates
  after BH formation.}
  \label{fig:bns.rho}
\end{figure*}

We find that turbulent viscosity has a much less obvious impact on the
evolution of the \ac{HMNS} than what could have been anticipated on the
basis of the idealized model in Sec.~\ref{sec:implementation}.  In the
first few milliseconds after merger, turbulent transport results in a
\emph{decrease} of the compactness, as can be seen from the maximum
density evolution (Fig.~\ref{fig:bns.rho}; left panel). Over longer
timescales, the behavior is non-linear. The \run{25} \ac{HMNS} is the
most compact remnant and collapses to a \ac{BH} ${\sim}17\ {\rm ms}$
after merger. The \run{5} remnant is only slightly less compact than
that of the reference simulation \run{0}.  \ac{BH} formation occurs at
${\sim}20\ {\rm ms}$ and ${\sim}22\ {\rm ms}$ after merger for the
\run{5} and \run{0} binaries, respectively. The \run{50} \ac{HMNS} is
the least compact and does not collapse to a \ac{BH} within our
simulation time. For the models that collapse within our simulation
time, we observe the formation of a massive (${\sim}0.1\ M_\odot$)
accretion disk (Fig.~\ref{fig:bns.rho}; right panel). As could have
been anticipated, the accretion rate is larger for simulations with
larger $\ell_{\rm mix}$.

\begin{figure*}
  \plottwo{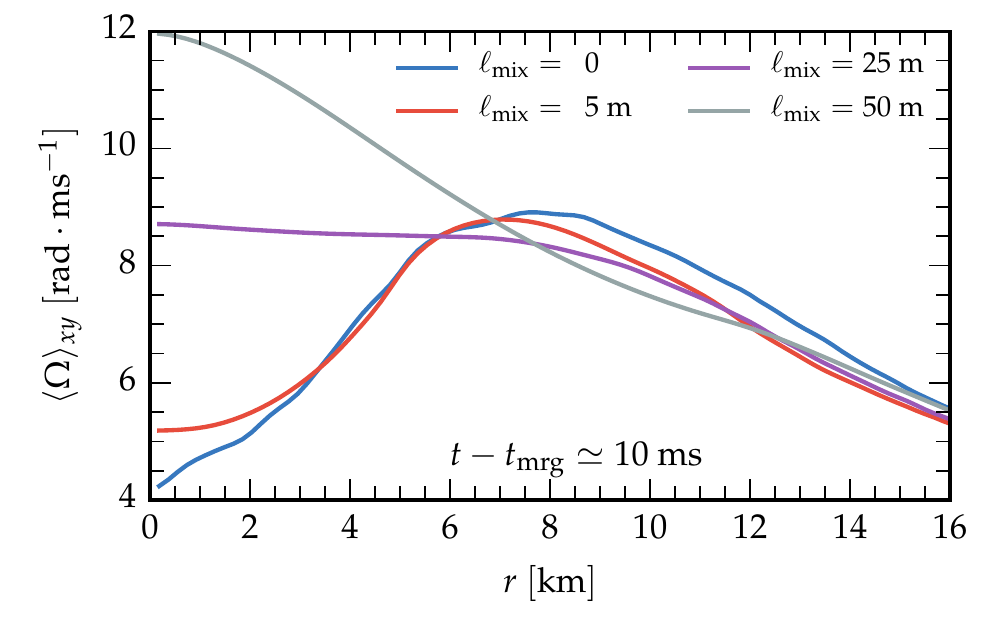}{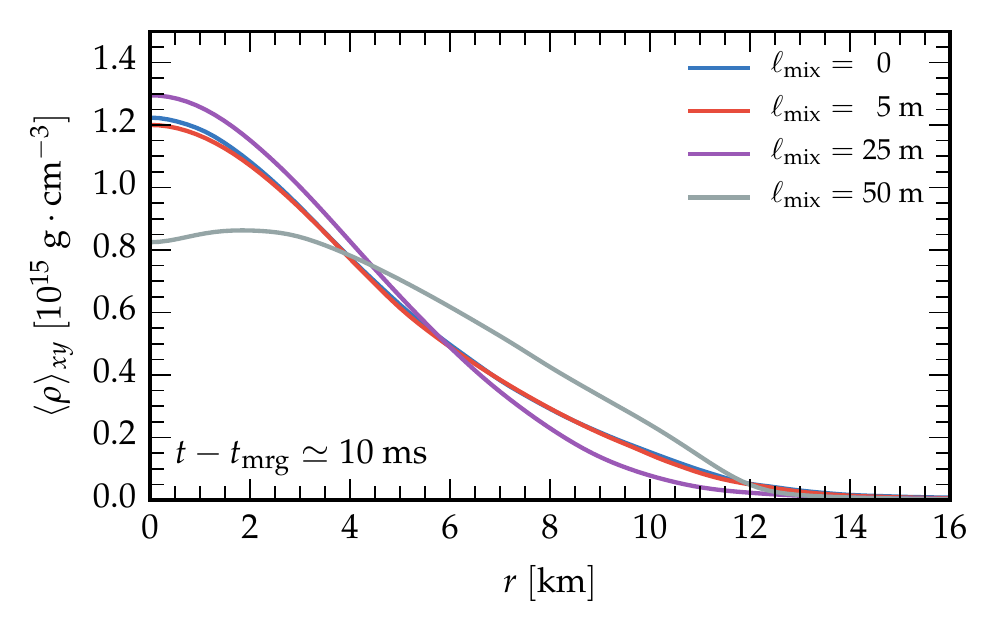}
  \plotone{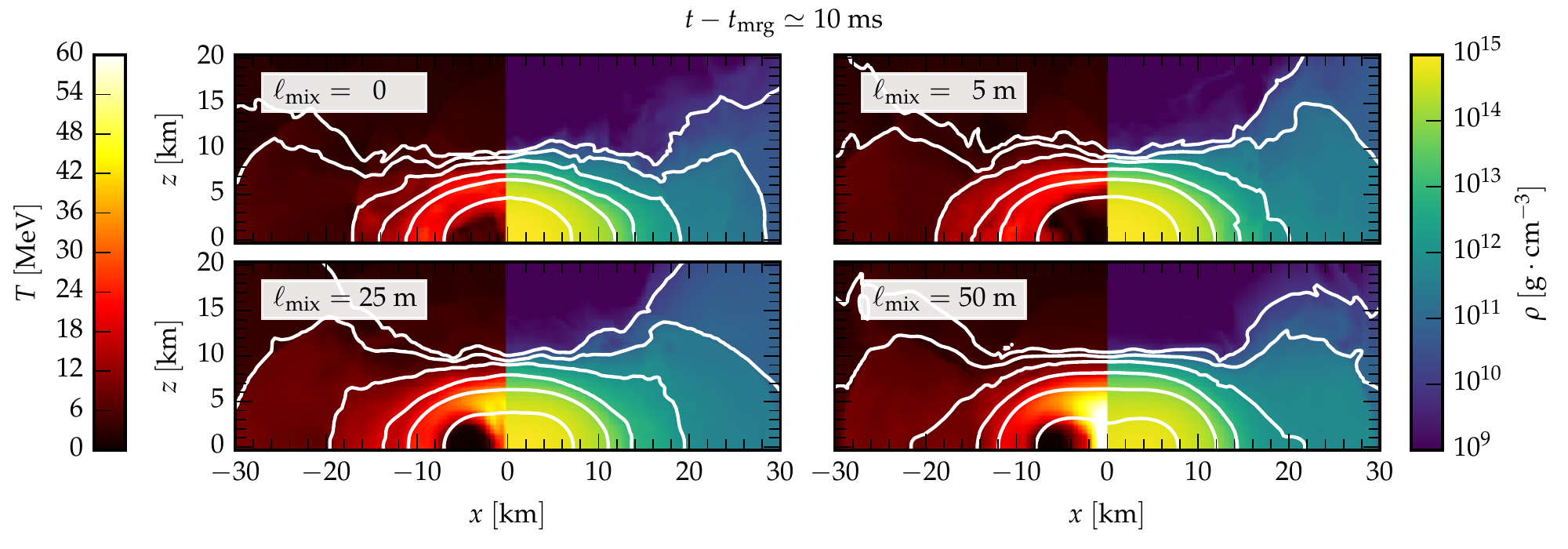}
  \caption{\emph{Upper panels:} angle-averaged angular velocity
  (\emph{left}) and density (\emph{right}) on the equatorial plane.
  \emph{Lower panels:} temperature and density in the meridional plane.
  All data are shown at ${\sim}10\ {\rm ms}$ after merger. The white
  contours in the lower panel are the isodensity contours for $\rho =
  10^{10}, 10^{11}, 10^{12}, 10^{13}, 10^{14}$, and $5\cdot 10^{14}\
  {\rm g}\cdot {\rm cm}^{-3}$. Turbulent dissipation leads to angular
  momentum transport and enhanced thermalization.}
  \label{fig:bns.hmns}
\end{figure*}

The reason for the different evolutions of the remnant can be understood
from the analysis of its internal structure (Fig.~\ref{fig:bns.hmns}).
The rotational profile established in the \ac{HMNS} after the initial,
very dynamical, phase is qualitatively different from that of
Eq.~\eqref{eq:jconst}, as also pointed out by \citet{shibata:2005ss,
kastaun:2016yaf, hanauske:2016gia, ciolfi:2017uak}. Consistently with
these previous studies, we find in the \run{0} simulation an \ac{HMNS}
composed of a slowly rotating core and a rotationally supported massive
envelope. As the mixing length increases, the structure of the \ac{HMNS}
is altered due to interplay between three competing effects. First,
angular momentum redistribution spins up the core, reducing its
compactness. Second, the loss of angular momentum from the massive
envelope results in a compression the \ac{HMNS}. Third, as more kinetic
energy is converted into thermal energy by turbulent dissipation, the
inner core becomes hotter and expands because of the increased pressure.
The interplay between these effects is complicated by the fact that the
angular momentum of the \ac{HMNS} is not conserved, but is radiated in
\acp{GW} at a rate proportional to that of the gravitational binding
energy \citep{bernuzzi:2015opx}. For this reason, as the \ac{HMNS}
contracts, it becomes more bound and at the same time it looses angular
momentum support.

The first and third effect are dominant at early times, so the effect of
turbulent viscosity is to monotonically reduce the compactness in the
first few milliseconds after merger. Later, all three effects become
important. At this stage, energy and angular momentum losses to \ac{GW}
play an important role. In the case of the \run{5} binary, the envelope
remains centrifugally supported (Fig.~\ref{fig:bns.hmns}; upper-left
panel), so the compactness is slightly decreased compared to the
reference run without turbulence dissipation. For the \run{25} binary,
the effect of turbulent transport is qualitatively similar to the
\run{5} binary at early times. Later, its envelope contracts causing the
growth of the central density (Fig.~\ref{fig:bns.rho}; left panel) and
early \ac{BH} formation. Finally, in the case of the \run{50} run the
thermal effect prevails; the hot spots formed in the contact layer at
the time of merger sink to the center and enhance the core temperature
to ${\sim}70\ {\rm MeV}$. The increased thermal support in the layers
with ${\sim} 5\cdot 10^{14}\cdot{\rm g}\cdot{\rm cm}^{-3}$ inflates the
\ac{HMNS}. The reduced compactness, in turn, results in a decrease of
the angular momentum loss due to \ac{GW} and prevents its collapse
within the simulation time.

\begin{figure*}
  \plottwo{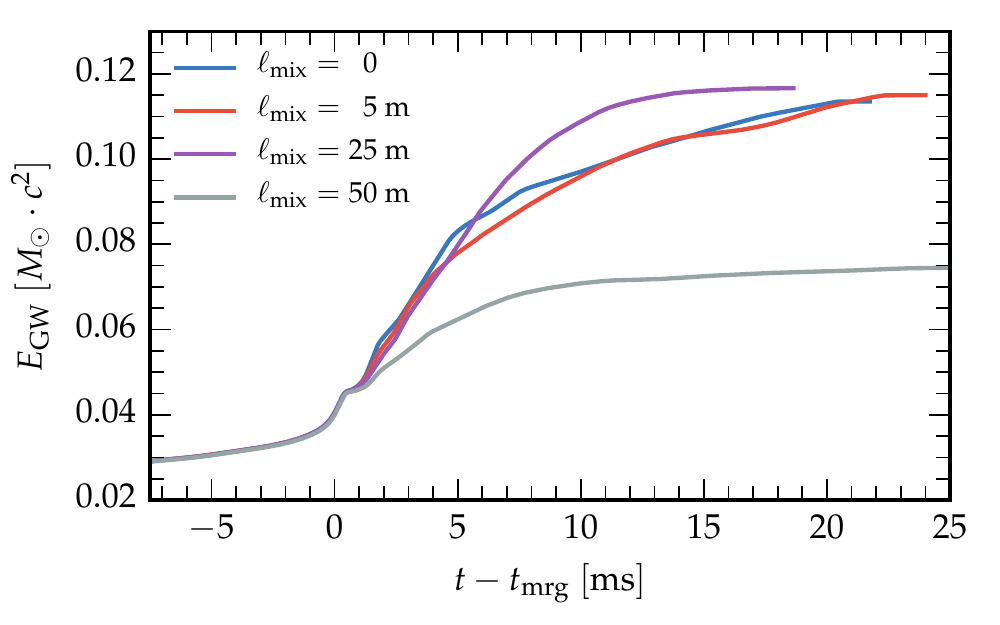}{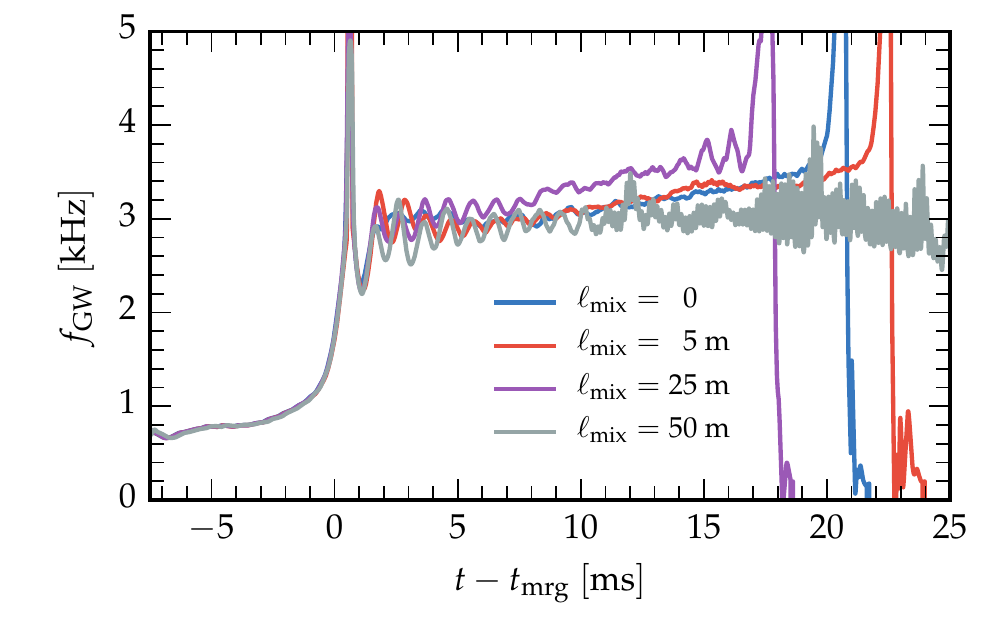}
  \caption{Total energy radiated in \ac{GW} (\emph{left panel}) and
  instantaneous \ac{GW} frequency (\emph{right panel}). The former is
  smoothed using a running average with a 0.1-ms window. Turbulent
  transport can influence the \ac{GW} luminosity starting from the
  early post-merger. The \ac{GW} instantaneous frequency is, instead,
  only weakly affected.}
  \label{fig:bns.gw}
\end{figure*}

The changes in the \ac{HMNS} structure are reflected in its
multimessenger emissions. The total energy radiated in \acp{GW}
(Fig.~\ref{fig:bns.gw}; left panel) is closely related to the rate of
increase of the \ac{HMNS} compactness. For this reason, at early times,
the amplitude of the signal slightly decreases with $\ell_{\rm mix}$,
while, over longer timescales, the behavior is non-monotonic. The
characteristic \ac{GW} frequency after merger (Fig.~\ref{fig:bns.gw};
right panel) is, instead, only weakly affected, with the exception of a
slight growth before \ac{BH} formation, which is a commonly observed
feature \citep[\eg][]{radice:2016rys}.

\begin{figure*}
  \plotone{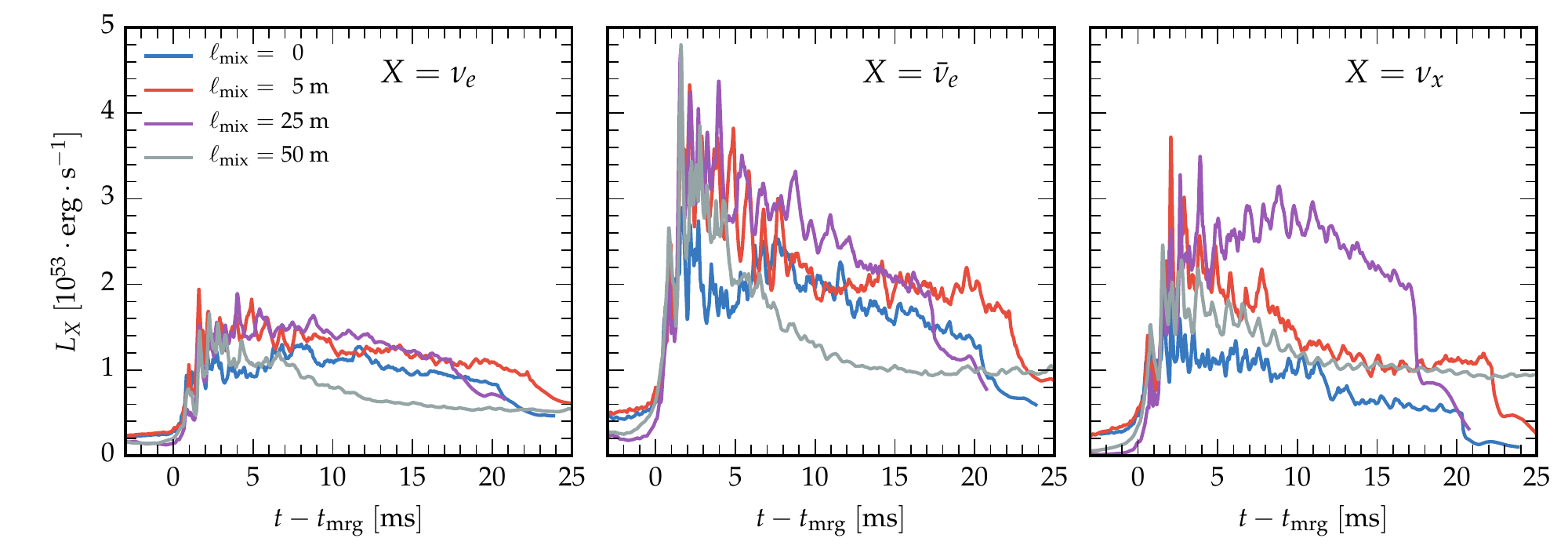}
  \caption{Electron (\emph{right panel}), anti-electron (\emph{middle
  panel}) and heavy-lepton (\emph{right panel}) neutrino luminosities.
  The increased temperature of the \ac{HMNS} due to turbulent
  dissipation leads to an increase in the neutrino luminosity for all
  species. This effect seems to be partially suppressed for the \run{50}
  simulation. The sudden drops in the emission for some of the
  simulations ${\sim}20$ after merger are due to \ac{BH} formation.}
  \label{fig:bns.nulum}
\end{figure*}

The neutrino emission (Fig.~\ref{fig:bns.nulum}) is also influenced by
the turbulent dissipation and the consequently higher temperatures in
the \ac{HMNS}. The luminosity of neutrinos of all flavours increases
with the mixing length parameter up to $\ell_{\rm mix} = 25\ {\rm m}$.
The luminosity of the \run{50} simulation is, however, smaller than that
of the \run{25} simulation. This is possibly because, in the \run{50}
\ac{HMNS}, the maximum of the temperature occurs at the center, while
for the other models it is off-centered \citep[see
also][]{kastaun:2016yaf}.

\section{Discussion and Conclusions} %
\label{sec:conclusions}
We have developed a new framework for the modeling of turbulence in
full-\ac{GR} simulations. Our approach is based on a relativistic
extension of the large-eddy simulation technique, which represents the
state-of-the-art for turbulence modeling in classical hydrodynamics
\citep{miesch:2015a}. Our method can naturally exploit turbulent
closures developed in Newtonian physics, is simple to implement, robust,
and stable.

As a first application, we have employed a turbulent viscosity closure
to study the effect of angular momentum transport and dissipation in
\ac{NS} mergers. We have performed, for the first time,
general-relativistic large-eddy simulations of merging \ac{NS} with
microphysical nuclear \ac{EOS} and neutrino cooling. We have found that
turbulence can modify the structure and collapse time of the merger
remnant. These, in turn, are reflected in the \ac{GW} and neutrino
emissions from the \ac{HMNS}. The accretion rate after \ac{BH} formation
is also affected.

The total energy radiated in \ac{GW} is the most affected quantity,
since it closely tracks the contraction of the \ac{HMNS} on its way to
the final collapse to \ac{BH}. In the presence of very efficient
turbulent transport, the effective viscosity might mask changes in the
compactness of the \ac{HMNS} that would otherwise be attributable to
changes in the high-density component of the \ac{EOS}
\citep{radice:2016rys}. This effect is, however, only modest for more
conservative choices of the turbulent mixing-length parameter. In these
cases, turbulence would not significantly affect the prospect of
detecting phase transitions in the core of the \ac{HMNS} using \ac{GW}
observations. However, a definitive statement will have to wait until
sufficiently resolved \ac{GRMHD} simulations are available to estimate
$\ell_{\rm mix}$.

We have also found that the post-merger \ac{GW} frequency is only weakly
affected by the effective turbulent viscosity. Thus, our results provide
an important validation of the several proposed methods relying on its
measure to constrain the \ac{EOS} of dense nuclear matter
\citep{bauswein:2011tp, takami:2014zpa, bernuzzi:2015rla}. Our results
also reaffirm the observation by \citet{bernuzzi:2015rla} that the
post-merger \ac{GW} peak-frequency is set at the time of merger.
Afterwards, the frequency stays close to constant and is largely
insensitive to the evolution of the \ac{HMNS}, with the exception of the
signature of \ac{BH} formation.

Finally, our results show that the neutrino signal is also influenced by
the turbulent dissipation of kinetic energy into heat. The increased
temperatures and luminosities, especially for the anti-electron
neutrinos, will influence the proton fraction in the outflows and might
have an effect on the resulting nucleosynthetic yields
\citep{wanajo:2014wha, metzger:2014ila, foucart:2016rxm}. Our results
strongly suggest that turbulent dissipation will have to be included in
the next generation of neutrino-radiation-hydrodynamics models of the
outflows from merging \acp{NS}.

Here, we presented a first application of the \ac{GRLES} method. In the
future, on the one hand, we will extend the present study to more binary
configurations and \ac{EOS}. On the other hand, work is already underway
to develop closures tuned with highly-resolved \ac{GRMHD} simulations
of \acp{HMNS}. Finally, we will extend \ac{GRLES} to \ac{GRMHD} and
couple it with a subgrid-scale dynamo model such as the one of
\citet{giacomazzo:2014qba}.

\begin{acknowledgments}
\acknowledgments
It is a pleasure to thank S.~Bernuzzi for the many stimulating
discussions on binary neutron star mergers. I also thank S.~Hild for the
ET-D noise curve data and A.~Burrows, P.~M\"osta, L.~Rezzolla, and
C.~D.~Ott for discussions. I gratefully acknowledge support from the
Schmidt Fellowship and the Sherman Fairchild Foundation. The simulations
were performed on Stampede NSF XSEDE (TG-PHY160025), and employed
computational resources provided by the TIGRESS high performance
computer center at Princeton University, which is jointly supported by
the Princeton Institute for Computational Science and Engineering
(PICSciE) and the Princeton University Office of Information Technology.
\end{acknowledgments}



\acrodef{ADM}{Arnowitt-Deser-Misner}
\acrodef{AMR}{adaptive mesh-refinement}
\acrodef{BH}{black hole}
\acrodef{BBH}{binary black-hole}
\acrodef{BHNS}{black-hole neutron-star}
\acrodef{BNS}{binary neutron star}
\acrodef{CCSN}{core-collapse supernova}
\acrodefplural{CCSN}[CCSNe]{core-collapse supernovae}
\acrodef{CMA}{consistent multi-fluid advection}
\acrodef{DG}{discontinuous Galerkin}
\acrodef{HMNS}{hypermassive neutron star}
\acrodef{EM}{electromagnetic}
\acrodef{ET}{Einstein Telescope}
\acrodef{EOB}{effective-one-body}
\acrodef{EOS}{equation of state}
\acrodefplural{EOS}[EOS]{equations of state}
\acrodef{FF}{fitting factor}
\acrodef{GR}{general-relativistic}
\acrodef{GRLES}{general-relativistic large-eddy simulation}
\acrodef{GRHD}{general-relativistic hydrodynamics}
\acrodef{GRMHD}{general-relativistic magnetohydrodynamics}
\acrodef{GW}{gravitational wave}
\acrodef{ILES}{implicit large-eddy simulations}
\acrodef{LIA}{linear interaction analysis}
\acrodef{LES}{large-eddy simulation}
\acrodefplural{LES}[LES]{large-eddy simulations}
\acrodef{MRI}{magnetorotational instability}
\acrodef{NR}{numerical relativity}
\acrodef{NS}{neutron star}
\acrodef{PNS}{protoneutron star}
\acrodef{SASI}{standing accretion shock instability}
\acrodef{SGRB}{short $\gamma$-ray burst}
\acrodef{SN}{supernova}
\acrodefplural{SN}[SNe]{supernovae}
\acrodef{SNR}{signal-to-noise ratio}

\end{document}